\documentclass[aps,pre,reprint,superscriptaddress]{revtex4-1}

\usepackage{graphics}
\usepackage[dvips]{graphicx}
\usepackage{subfig}
\usepackage{amsmath}
\usepackage{amsfonts}
\usepackage{hyperref}
\usepackage[belowskip=-10pt]{caption}

\begin{document}

\title{Memory order decomposition of symbolic sequences}

\date{\today}

\author{Unai Alvarez-Rodriguez}
\affiliation{Basque Centre for Climate Change (BC3), 48940 Leioa, Spain}
\affiliation{Data Analytics Group, University of Zurich, CH-8006 Zurich, Switzerland}
\affiliation{School of Mathematical Sciences, Queen Mary University of London, London E1 4NS, United Kingdom}

\author{Vito Latora}
\affiliation{School of Mathematical Sciences, Queen Mary University of London, London E1 4NS, United Kingdom}
\affiliation{Dipartimento di Fisica ed Astronomia, Universit{\`a} di Catania and INFN, 95123 Catania, Italy}
\affiliation{The Alan Turing Institute, The British Library, London NW1 2DB, United Kingdom}

\begin{abstract}
We introduce a general method for the study of memory in symbolic sequences based on higher-order Markov analysis. 
The Markov process that best represents a sequence is expressed as a mixture of matrices of minimal orders, enabling the definition of the so-called memory profile, which unambiguously reflects the true order of correlations.
The method is validated by recovering the memory profiles of tunable synthetic sequences. 
Finally, we scan real data and showcase with practical examples how our protocol can be used to extract relevant stochastic properties of symbolic sequences.
\end{abstract}

\maketitle

\section{Introduction}
Symbolic sequences are ubiquitous in many domains of science.
For instance, we use sequences of symbols to encode the sounds that constitute our different languages, to disentangle the complexity of DNA molecules responsible for our genetic information,
and also to characterize temporal evolution of physical systems.
In this context, memory can provide key information about a sequence and the process generating it, as it represents the distance between causally related elements in the sequence.
The Markov chain formalism \cite{time1,markov1,markov2,markov3}, allows for an approximation of the generating process by means of a maximum likelihood estimator of a given memory or order.
The problem of extracting the order of a generating process has been addressed by means of different information criteria \cite{ic1,ic2,99buiatti,ic3,ic4,ic5,ic6,ic7,ic8,ic9,ic10,ic11} 
(belonging to the more general field of model selection \cite{ms1,ms2}), which provide estimates of the maximal order as a function of the likelihood and the number of parameters involved in the model. 

Various higher-order models have been proposed as a formal way to analyze memory in a complex system \cite{12holme,13pfitzner,14rosvall,14scholtes,18lacasa,19lambiotte,19williams}.
These models allow one to go from a time-aggregated perspective to a dynamics that respects the time ordering of interactions.
Recently, there is a growing interest for combining the statistics of different orders into a single model; see \cite{17scholtes,20gote} for multi order models,
and see \cite{17melnik} for a decomposition of transition probabilities in terms of generalized memory functions.
However, a general and analytic framework for understanding the nested nature of memory is still missing. 

In this article we address this gap by introducing the concept of memory profile of a stochastic process,
and designing an algorithm that captures the length specific correlations present in the temporal evolution of a system.
Our method decomposes a Markov process into a convex sum of stochastic matrices, where the memory profile arises naturally as the set of coefficients of the subprocesses.
The algorithm detects which of the correlations at a given length are spurious, by explaining them as a combination of subprocesses of lower orders.
We finally validate the method on synthetic sequences and we illustrate how it works in practice to extract the memory profiles of real data coming respectively from literary texts, biology and deterministic chaotic systems.

Let us start with an empirically observed sequence $S=(s_1, s_2, \ldots, s_L)$ of length $L$, where the generic symbol $s_i$, with $i=1,2,\ldots,L$, is selected from an alphabet ${\cal A} = \{a_1,a_2, \ldots, a_A\}$ of $A$ characters.
We assume $S$ has been created from an unknown Markov process $\mathcal{Q}$ that we call source, which can be expressed as a stochastic matrix $Q$.
In order to study the statistical properties of $\mathcal{Q}$ through $S$, let us define as $x^m = (x_1 , \ldots, x_m)$ an ordered string of $m$ elements of $\mathcal{A}$.
Let us also denote as $s^m_i$ the string of length $m$ terminating at position $i$ in $S$. 
For simplicity we will refer to strings of length $1$, single elements, with a reduced notation, using $s_i$ instead of $s^1_i$.
Each of the $s^m_i$ corresponds to a unique string $x^m$, 
while the probability of finding a particular $x^m$ in $S$ is given by $p(x^m)=\frac{f(x^m)}{L-m+1}$, where $f(x^m)$ denotes the number of appearances of $x^m$ in $S$. 
Assuming that sequence $S$ is generated by $Q$ of order $m$, or $Q^m$, the {\it transition probabilities} can be estimated as
\begin{equation}  
\pi (x_m | x_1, \ldots , x_{m-1}) = \frac{f(x^m)}{f^-(x^{m-1})},
\label{tp}
\end{equation}
where the reduced frequencies $f^-(x^m)$, are equal to $f(x^m)$, if $x^m \neq s^m_L$, or to  $f(x^m)-1$ if $x^m = s^m_L$. 

The transition probabilities can now be organized in an $A \times A^{m-1}$ transition matrix, $T^m$, which, for a given order $m$,
contains the probabilities $\pi (x_m | x_1, \ldots , x_{m-1})$ in Eq.~(\ref{tp}) to get any of the $A$ symbols after each of the possible ordered combinations of $A^{m-1}$ symbols.
See the Appendix, Sec. \ref{ap:mn} for an example of the matrix notation.

A measure of how good a model $T^m$ of order $m>0$ is for the observed sequence $S$ can be obtained by the likelihood function 
\begin{equation}
\ell (T^m,S)= p(s_1) \prod^{m-1}_{i=2}\pi(s_i|s_{i-1}^{i-1}) \prod_{i=m}^L \pi(s_i|s_{i-1}^{m-1})
\end{equation}
where the transition probabilities are computed with the $x^{m}$ associated to each $s^{m}_i$. The likelihood is a non decreasing function of $m$, and estimating the maximal order $M_{Q}$ of $Q$ as the value of $m$ at which $\ell$ is maximal will lead to overfitting. To cope with that we extract $M_Q$ via the Akaike information criterion (AIC), which formalizes the intuitive idea of a trade-off between the number of parameters and the performance of a model~\cite{ic1}. 

\section{Memory order decomposition}
\subsection{Nested models}
If a sequence is perfectly described by a model of order $m$, the transition probabilities at $m+1$ should return the same description.
Let us address this problem by defining $T^{m[+]1}$, a prediction of how $T^{m+1}$ should be, assuming $T^m=Q$.
In practice, a transition matrix $T^m$ from an alphabet $\mathcal A$ is extended by taking the tensor product  $T^{m[+]j} = {\bf{1}} \otimes T^{m}$ with a row vector ${\bf{1}}$ of length $A^{j}$ with all the components equal to 1. We can now compute the distance $d$ between $T^{m+1}$ and $T^{m[+]1}$ to test the robustness of the $T^m = Q$ hypothesis. We will be using $d=1-\sigma$, where $\sigma$ measures the overlap between two discrete probability vectors $u$ and $v$ of dimension $D$, and it is expressed as $\sigma(u,v)=\sum^{D}_{i} \min (u_i, v_i)$. By construction $\sigma(u,v)$ is bounded to the $[0,1]$ interval, with $\sigma=1$ if $u=v$. This definition of $d$ is equivalent to a normalized vector distance, as we show in the Appendix, Sec. \ref{ap:sd}. In our case we deal with matrices, so we will be calculating the statistical distance for each of the columns, and afterwards weighting the contribution of columns $\alpha$, with $\alpha=0,\ldots, A^{m-1}-1$, by the probability $p(x^{m-1}_\alpha)$ of finding the string corresponding to column $\alpha$ in $S$. 
The final expression is:
\begin{equation}
\sigma(T^{m+1},T^{m[+]1})=\sum^{A^{m}-1}_{\alpha=0} p(x^{m}_\alpha) \sum^{A-1}_{\beta=0} \min (T^{m+1}_{\beta\alpha}, T^{m[+]1}_{\beta\alpha}),
\label{eqsigma}  
\end{equation}

\subsection{Decomposition}
Building on the same idea, it is possible that a matrix $T^m$ has a non-zero overlap with its predecessors $T^{m-1}, \ldots, T^{0}$, implying that in the procedure of generating $S$ not all of the new elements depend on the previous $M_Q$ elements; some of them could have required much shorter strings, i.e., a shorter memory. In this case, the very same idea of the true order of $\mathcal{Q}$ would be misleading. We will now show that it is possible to extract the memory profile, i.e., the relevance of each order $m \le M_T$ in $T^{M_T}$, by adopting a matrix decomposition procedure as follows. In general, any column-stochastic matrix, such as the transition matrix $T^{M_T}$, can be decomposed as a linear combination of deterministic processes of different orders, i.e., column-stochastic Boolean matrices $C^m_i$ of dimension $A \times A^{m-1}$ as
\begin{equation}
T^M=c^{0}_0 C^{0[+]M_T -1}_0+\sum^{M_T}_{m=1} \sum_{i=1}^{\mathcal{C}^m} c^m_i C^{m[+](M_T-m)}_i ,
\label{dek}
\end{equation}
where $\mathcal{C}^m$ stands for the number of deterministic processes at each order, the coefficients $c^m_i$ are real numbers weighting the different contributions, and the $m=0$ process corresponds to a uniform model. The latter assigns an equal probability of $1/A$ to all the symbols in $\mathcal{A}$, and is considered separately from the other processes since $C^{0}_0$ is not Boolean. In order to visualize the total contribution of each order, we define  the {\it memory profile} of the transition matrix $T^{M_T}$ as the vector ${\bf t}$ whose components $t_m$, with $m=0,\ldots, M_T$, are given by $t_m = \sum_{i}^{\mathcal{C}_m} c^m_i $. Conversely, we say that $q_m$ represents the memory profile of the original process $\mathcal{Q}$. The particular form of the deterministic matrices $C^m_i$ allows a one-to-one correspondence  with natural numbers. In fact each of the columns of any of our $C^m_i$ contains a single nonzero element which is equal to 1. The position of this element can be associated to a term in a power expansion base $A$, where the row accounts for the coefficient and the column for the power. If $C^m_i$ has elements $e_{\alpha\beta}$, the associated number $n^m_i$ is $n^m_i=\sum^{A-1}_{\alpha=0} \sum^{A^{m-1}-1}_{\beta=0} e_{\alpha\beta} \alpha A^{\beta}$, while $n^{0}_{0}=0$ for the uniform model. Index $m$ in $n^m_i$ is necessary to avoid redundancies between processes at different orders with the same associated number (see the Appendix, Sec. \ref{ap:nl}).

Our goal for the mixture in Eq. \eqref{dek} is to have $c^m_i=0$ for all the matrices that are the extension of a lower order matrix, i.e., reducible matrices. The standard procedure for identifying these matrices is to test whether they correspond to the tensor product of a lower-order matrix. Alternatively the mapping to natural numbers introduced above allows one to simplify the problem, as the natural number $n^m_i$ associated to a process inherits its order properties. It is then enough to check whether $n^m_i$ is divisible by a given number to prove that $C^m_i$ has true order $m$. Let $n^m_i$ be the number associated to a given process $C^m_i$ and let $C^{m[+]1}_i$ and $n^{m[+]1}_i$ be its extension to the next order, and the number associated with it. We have:
\begin{equation}
n^{m[+]1}_i= n^{m}_i \sum^{A-1}_{\alpha=0} A^{\alpha A^{m-1}} = n^{m}_i \frac{A^{A^{m}}-1}{A^{A^{m-1}}-1}.
\label{natex}
\end{equation}
This formula provides a simple reduction mechanism: a given number $n^m_i$ has a true order $m$ if it is not divisible by $\frac{A^{A^{m-1}}-1}{A^{A^{m-2}}-1}$, otherwise it can be reduced. This check is then repeated until the number is found to be not divisible. The order in which the reduction process terminates is the true order of the process associated with this number (see the Appendix, Sec. \ref{ap:nre}).

\subsection{Algorithm}
The decomposition algorithm we propose here consists of an iterative procedure that, at each step, identifies the process with the maximal coefficient and removes it from the matrix to be decomposed. The transition to a higher order is produced after ensuring that no more processes can be added to the decomposition. See the Appendix, Sec. \ref{ap:da} for a fully detailed example of the algorithm.

The procedure is equivalent for each step, so let us suppose the matrix we want to decompose is $T^{M_T}$ which can represent the original transition matrix or any of its intermediate steps of decomposition. Let us also suppose that we are currently exploring the matrices at a generic order $m$. The first step is to create a reduced matrix from $T^{M_T}$, with the dimensions of the matrices at order $m$, $A \times A^{m-1}$. When $T^{M_T}$ is reduced to order $m$, each of the elements of the reduced matrix $R^m$ is fed with the elements of $T^{M_T}$ that correspond to its extension. The specific $R^m$ we are looking for, is the one where each matrix element is the minimal of all the elements in $T^{M_T}$ that correspond to the tensor product extension.
\begin{equation}
R^{m}_{ij} = \left\{\begin{array}{l} \min_{\alpha\beta} T^{M_T}_{\alpha\beta}  \hspace{0.2cm} \textrm{if} \hspace{0.2cm} m=0\\\min_\beta \{ T^{M_T}_{i\beta}| \beta \equiv j \mod A^{m-1} \} \hspace{0.2cm} \textrm{if} \hspace{0.2cm} m\neq 0\end{array}\right.
\label{eqRm}
\end{equation}

The second step is to select the matrix to be incorporated into the decomposition. This is done by finding the maximum in each of the columns of $R^m$. The matrix will be the one whose non-zero elements are located in the position of the maximum values. If there is more than one maximum, there is not a unique possible matrix, and we say that the process is degenerate. 

The third step is to detect the coefficient, $c^m_i$, of the process we have just found, $C^m_i$. The idea is that the enlarged form of the matrix, weighted with its corresponding coefficient, $c^m_i C^{m[+](M_T-m)}_i$, is subtracted from $T^{M_T}$. Therefore, in order to have the maximum of the non-negative outcomes, the coefficient has to be the minimum of the set of maximum column values in $R^m$. 
\begin{equation}
c^m_i = \min_\beta \max_\alpha  R^m_{\alpha\beta}.
\end{equation} 

As anticipated, the fourth step is to remove  $c^m_i C^{m[+]M_T-m}_i$ from $T^{M_T}$. The result of this process is a new $T^{M_T}$, which is the output of the current cycle and the input of the following one. The next cycle should repeat the search in $m$, unless the just found coefficient is $c^m_i=0$. This would mean that no more processes are compatible with $T^{M_T}$ at $m$. If the previous condition is true, the value of $m$ has to be updated to $m+1$. 

\subsection{Validation}
We carry out a systematic validation procedure on ensembles of synthetic sequences with different alphabets, maximal orders, and lengths. We have used two indicators ($v_1$,$v_2$), each of them a real number in $[0,1]$, to evaluate the performance of our method: $v_1$ accounts for the success of the AIC in retrieving the maximal order of $M_Q$, and $v_2$ measures the overlap $\sigma(q_m,t_m)$ between $t_m$ and $q_m$. Here, a sequence is generated by randomly constructing a column-stochastic matrix $Q^{M_Q}$ for each ($L$, $M_Q$, $A$) triplet. The output indicators $v_i$ are averaged over $100$ realizations of different experiments with the same ($L$, $M_Q$, $A$) values. Therefore, $v_1$ is the fraction of times $M_T=M_Q$ and $v_2$ is the average $\sigma(\mathbf{q},\mathbf{t})$, where $\mathbf{q}$ is the real memory profile.

\begin{figure}[htb]
\captionsetup[subfigure]{labelformat=empty}
\subfloat[]{\includegraphics[width=0.12\textwidth]{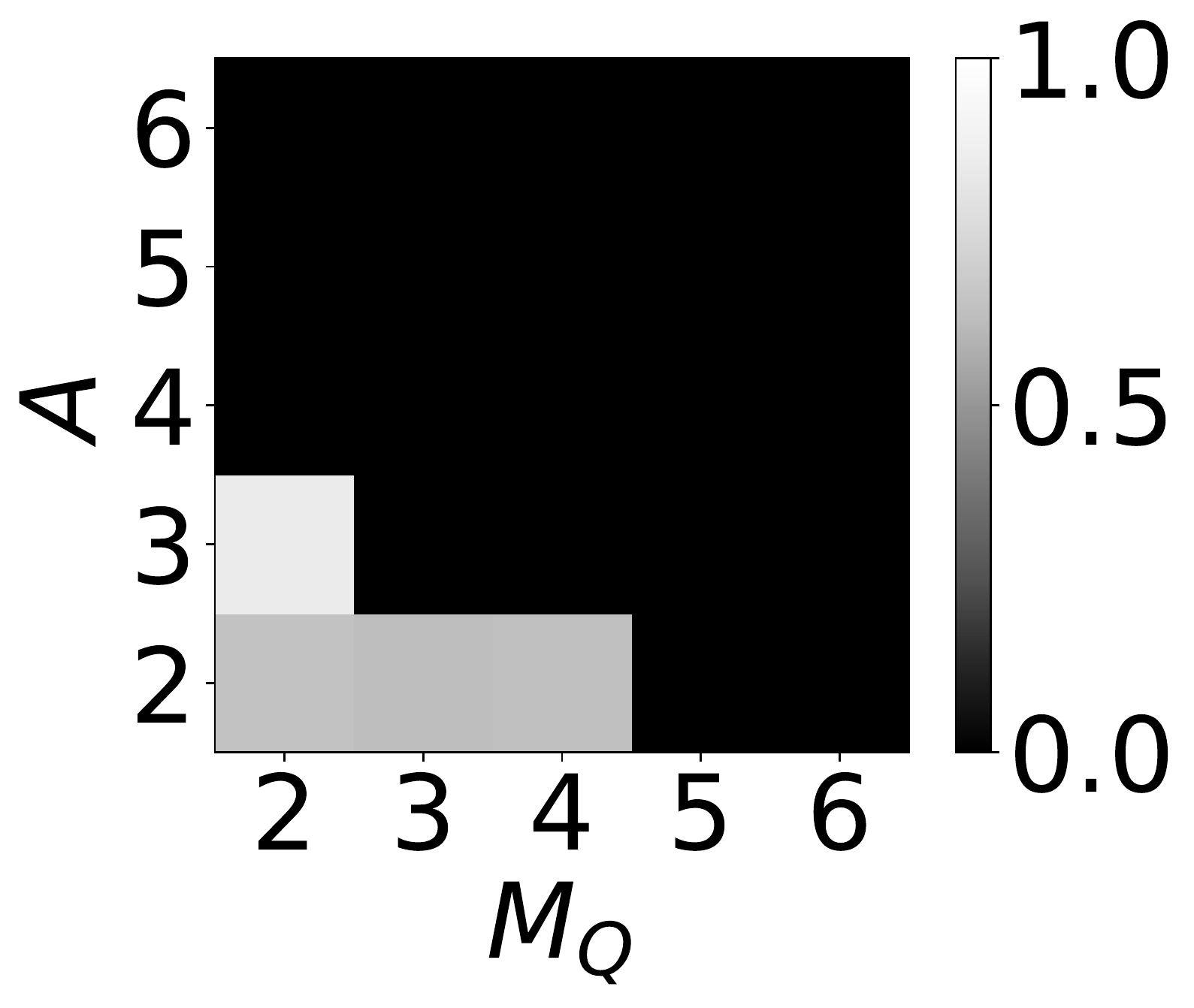}}
\subfloat[]{\includegraphics[width=0.12\textwidth]{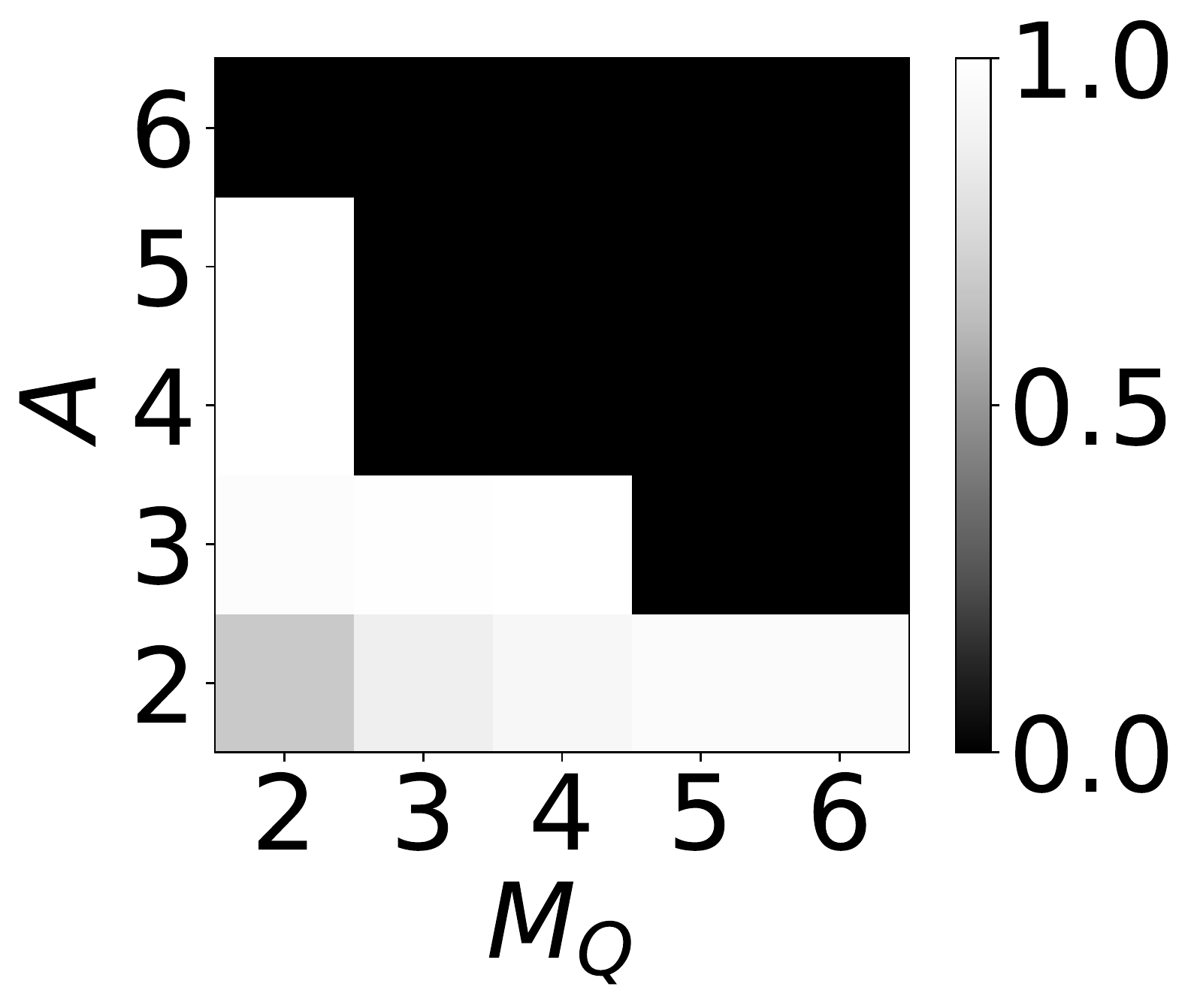}}
\subfloat[]{\includegraphics[width=0.12\textwidth]{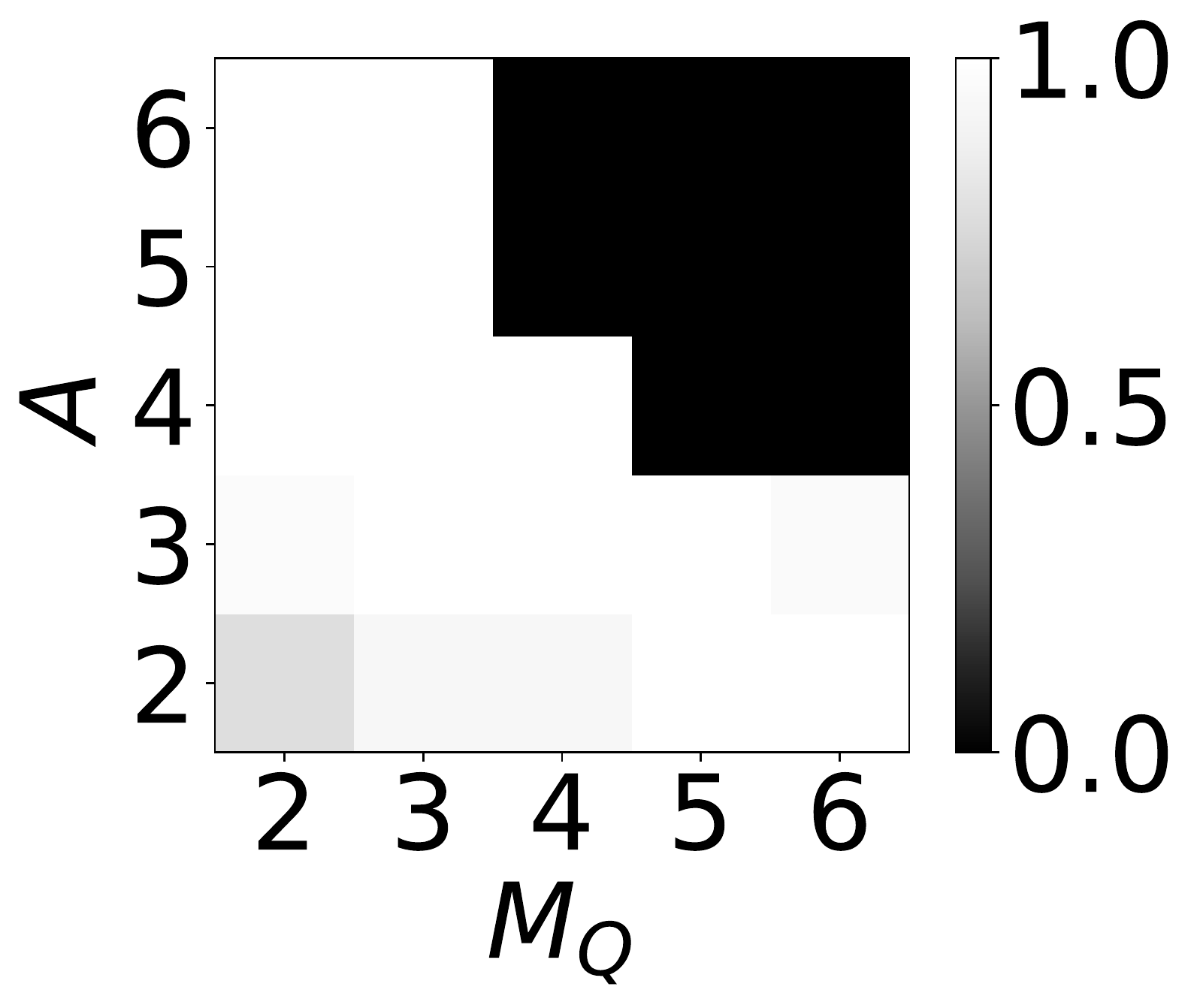}}
\subfloat[]{\includegraphics[width=0.12\textwidth]{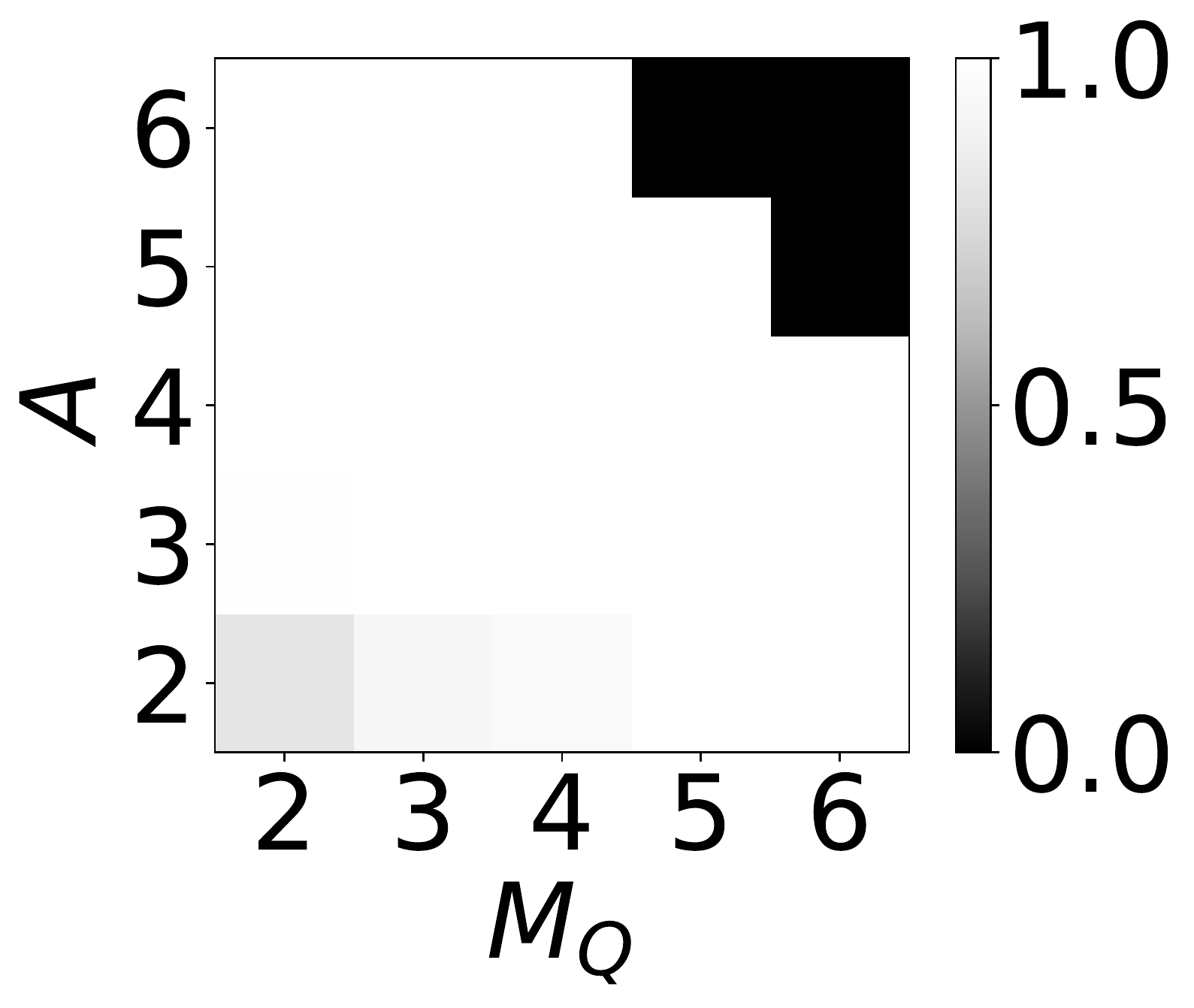}}\\[-5.75ex]

\subfloat[$L=10^2$]{\includegraphics[width=0.12\textwidth]{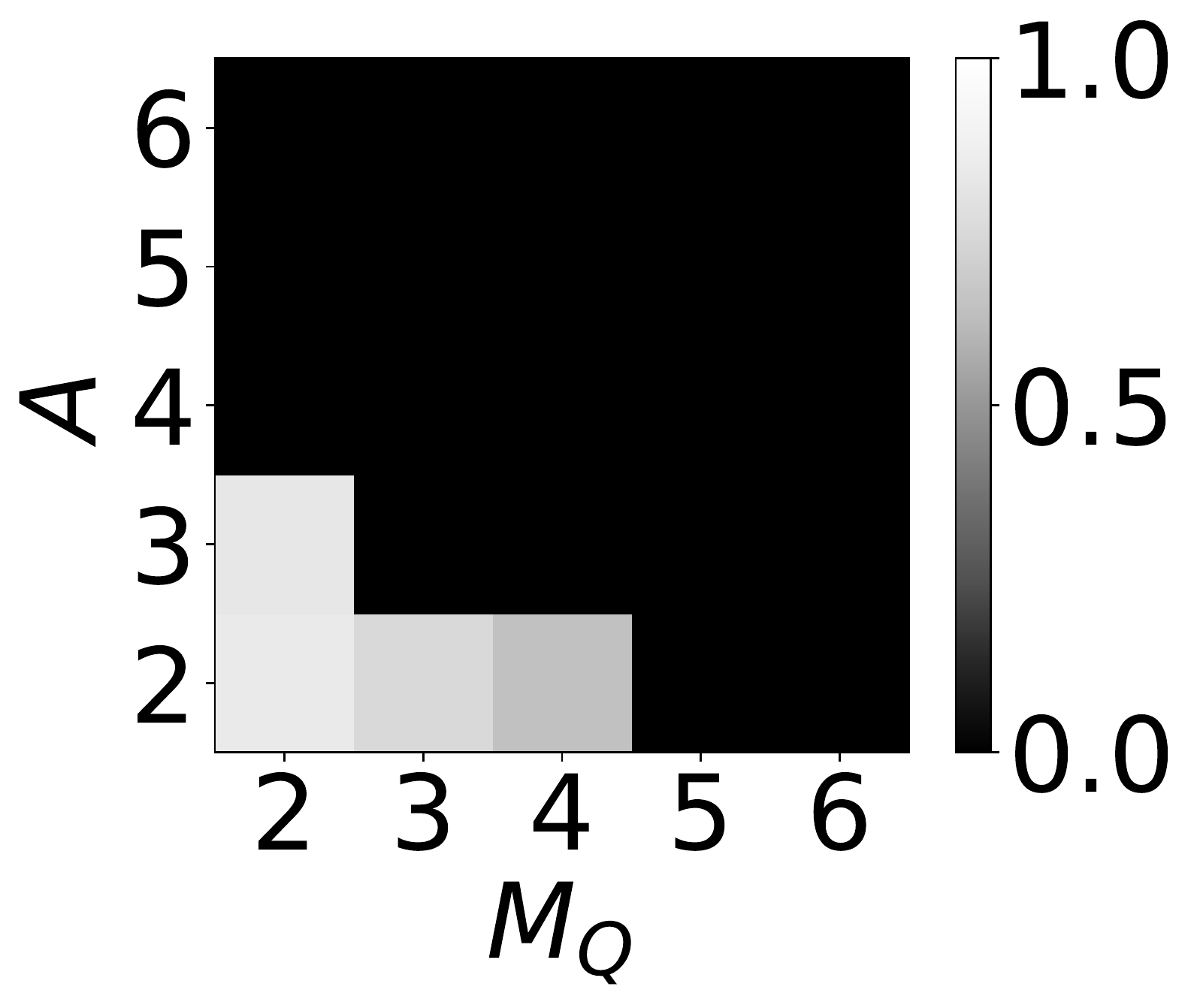}}
\subfloat[$L=10^3$]{\includegraphics[width=0.12\textwidth]{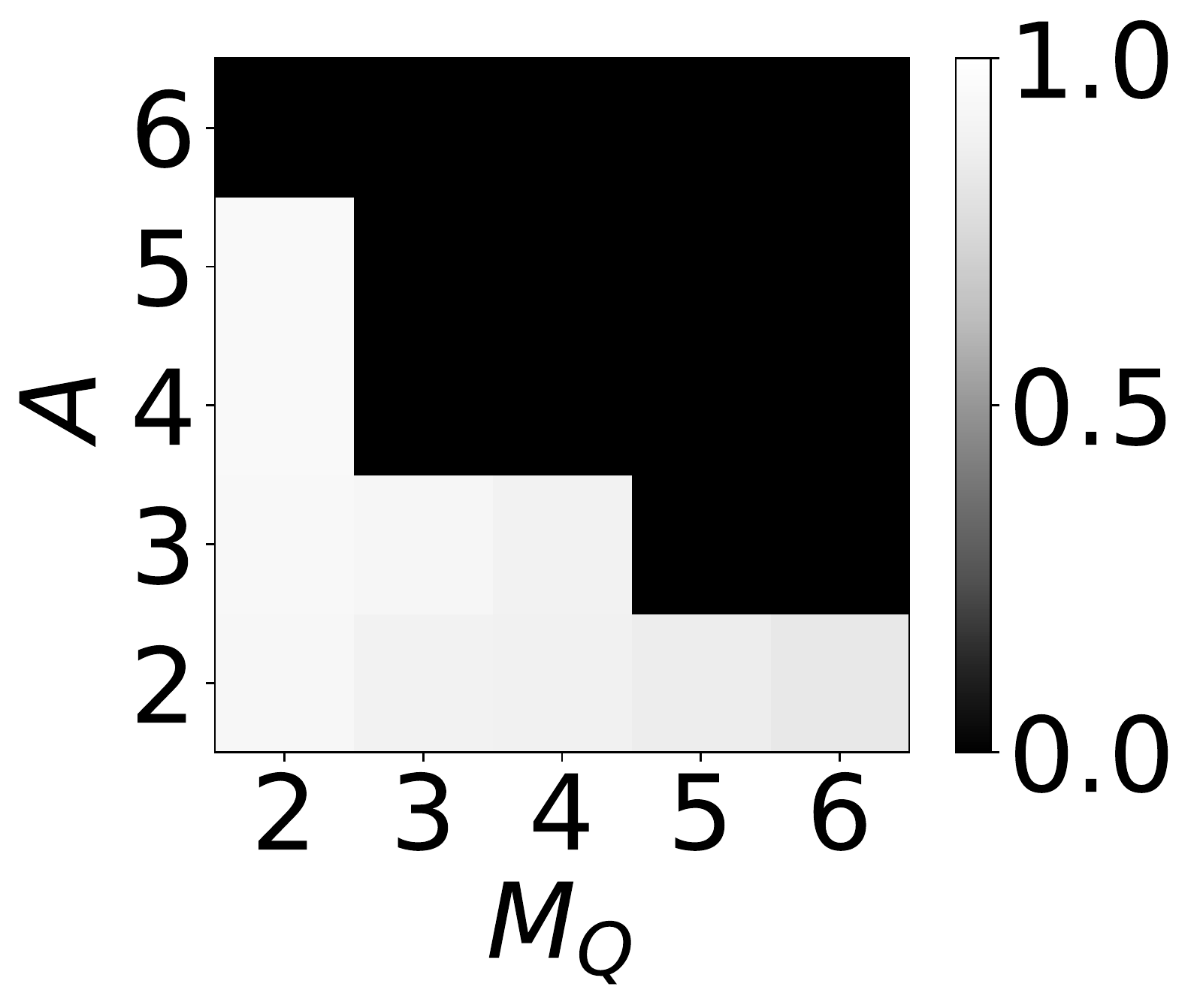}}
\subfloat[$L=10^4$]{\includegraphics[width=0.12\textwidth]{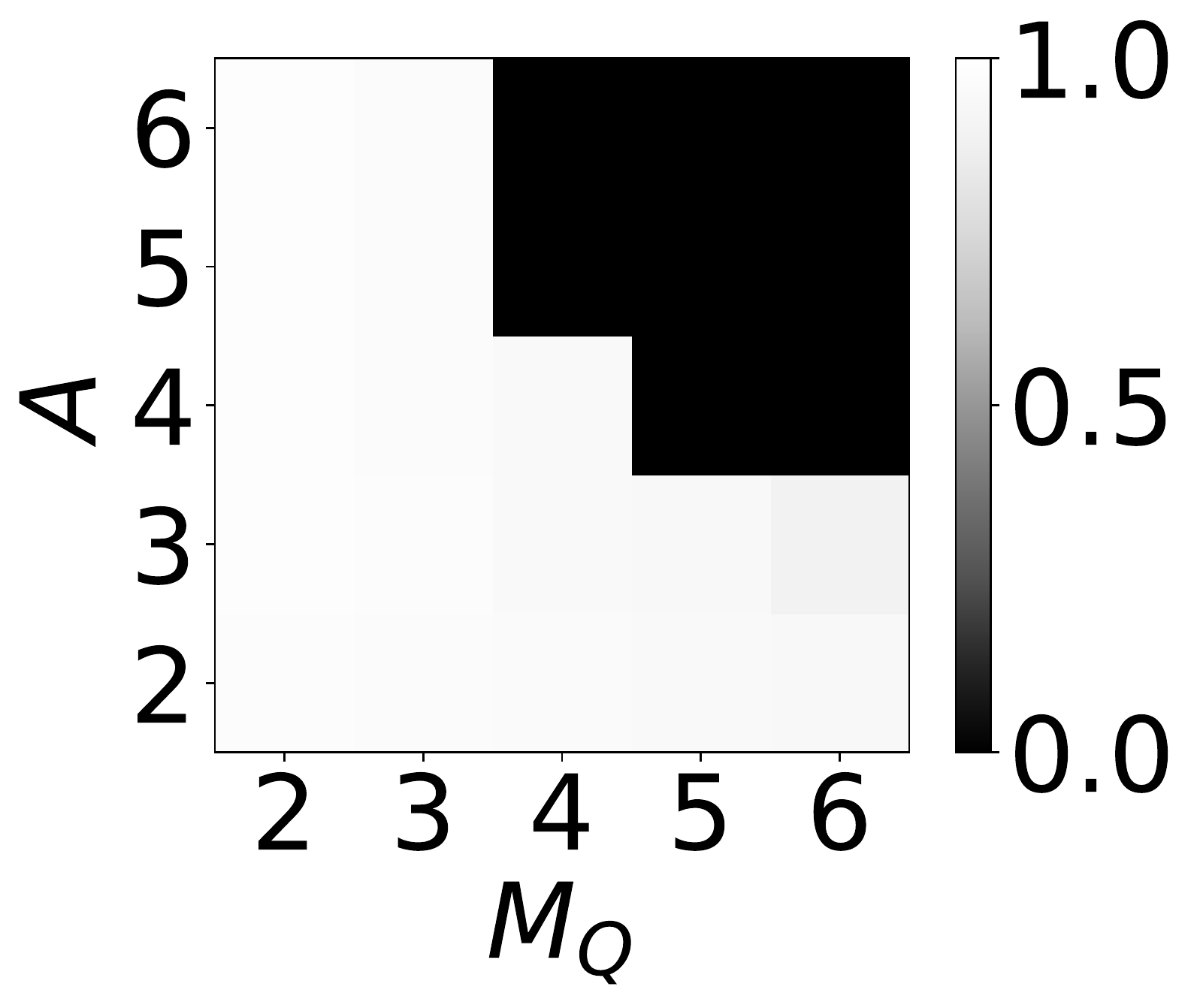}}
\subfloat[$L=10^5$]{\includegraphics[width=0.12\textwidth]{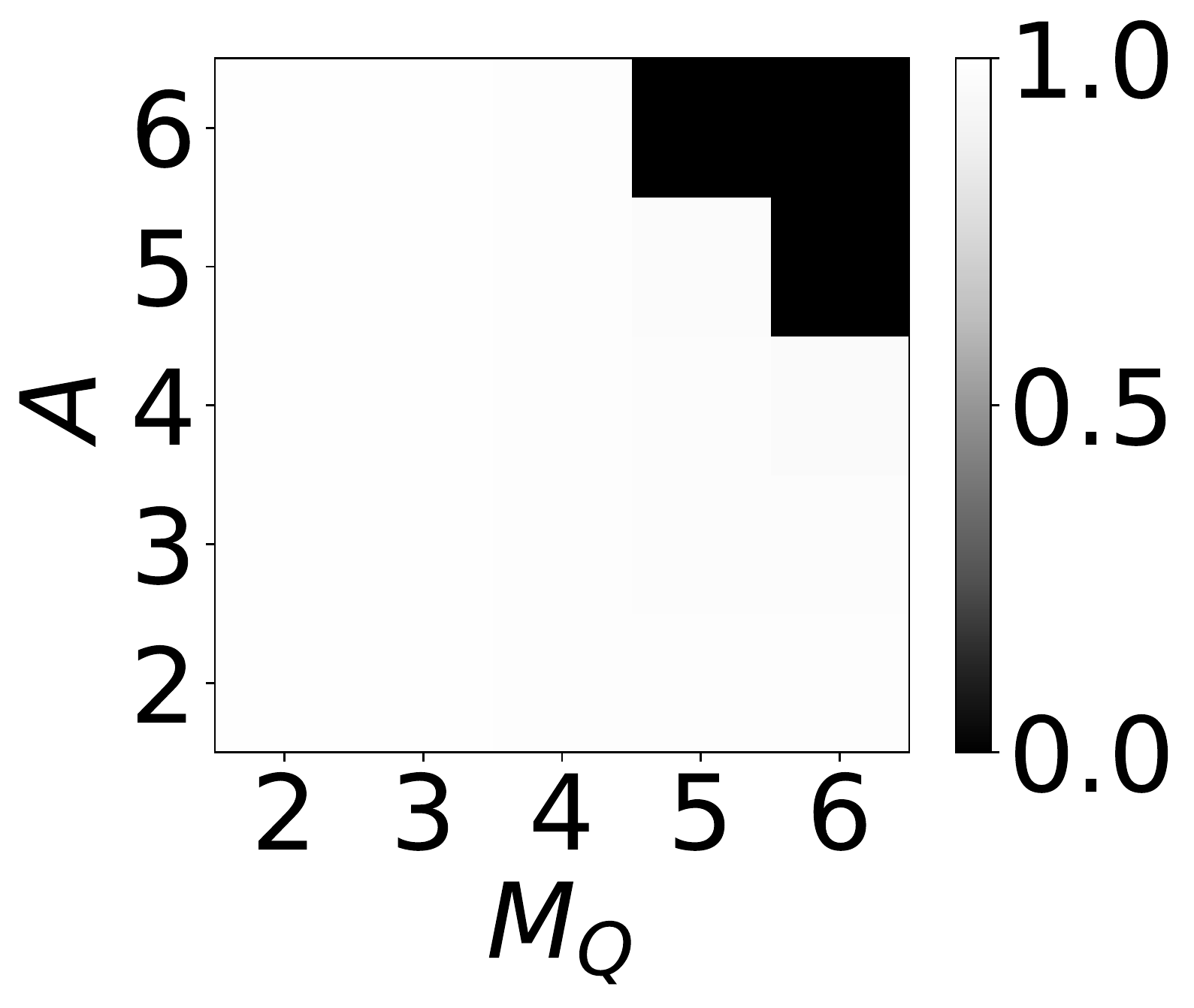}}
\caption{Validation of the algorithm on synthetic sequences with different values of length $L$, maximal order $M_Q$, and number of symbols $A$. The rows report, respectively, the computed values of $v_1$ and $v_2$ with a color code, while different columns refer to sequences with different values of $L$. Results are obtained as averages over $100$ different realizations.}
\label{syn}
\end{figure}

The results are shown in Fig.~\ref{syn}. The rows refer, respectively, to $v_1$, $v_2$. The color of each cell denotes the average values of $v_i$ as a function of $M_Q$ and $A$, and the columns account for four different sequence lengths. Notice from the figure that the reliability of the order detection is affected by the length of the sequence and that of the alphabet, in the sense that we impose an upper bound in $m$ that guarantees a minimal frequency for the strings of $T^{m+1}$. See the Appendix, Sec. \ref{ap:lfs} for the specific details of our treatment of low frequency strings. Below such threshold the behavior of the algorithm is satisfactory even for small values of $L$. It is noteworthy to mention that the errors in small $M_Q$ at $v_1$ are compensated in $v_2$. In other words, even in the cases in which the AIC fails the complete algorithm succeeds in extracting the memory profile.

\section{Application to real sequences}
We have extracted the memory profile of sequences from biology, literary texts, and chaotic systems. We have selected these datasets because they correspond to alphabets that we have tested with synthetic sequences, and because each of these examples showcases a new feature of sequence analysis enabled by our protocol: the true memory allocation across different orders, the non-trivial ranking of subprocesses, and the finiteness of the number of subprocesses involved in a higher-order Markov chain decomposition. In this sense, our goal here is not to address domain-specific questions. Again, we impose an upper bound to the highest memory order $m$ to ensure that the correlations that we find are not an effect of the finiteness of the data (see Appendix, Sec. \ref{ap:lfs}).

\begin{figure}[htb]
\captionsetup[subfigure]{labelformat=empty}
\subfloat[]{\includegraphics[width=0.16\textwidth]{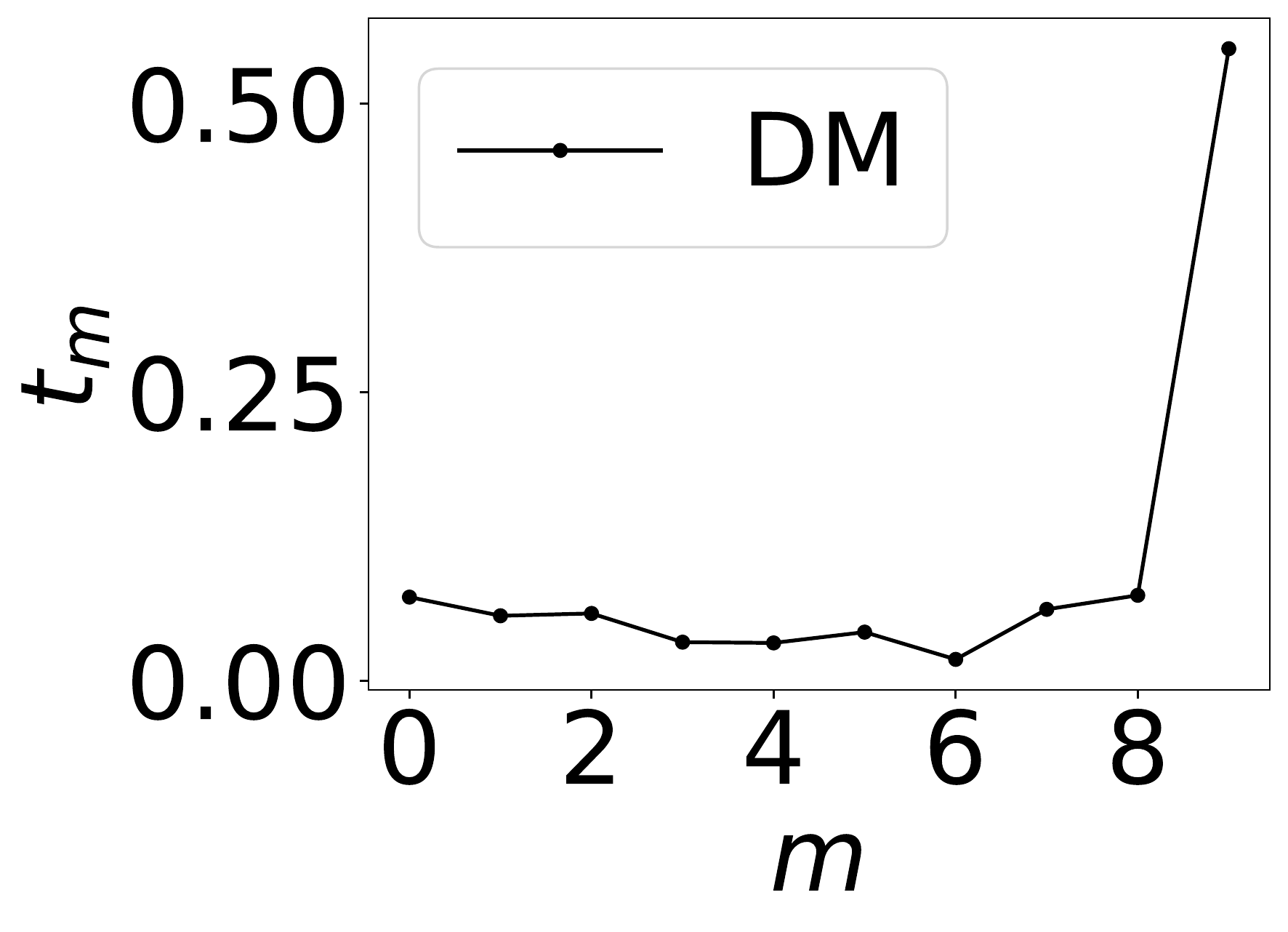}}
\subfloat[]{\includegraphics[width=0.16\textwidth]{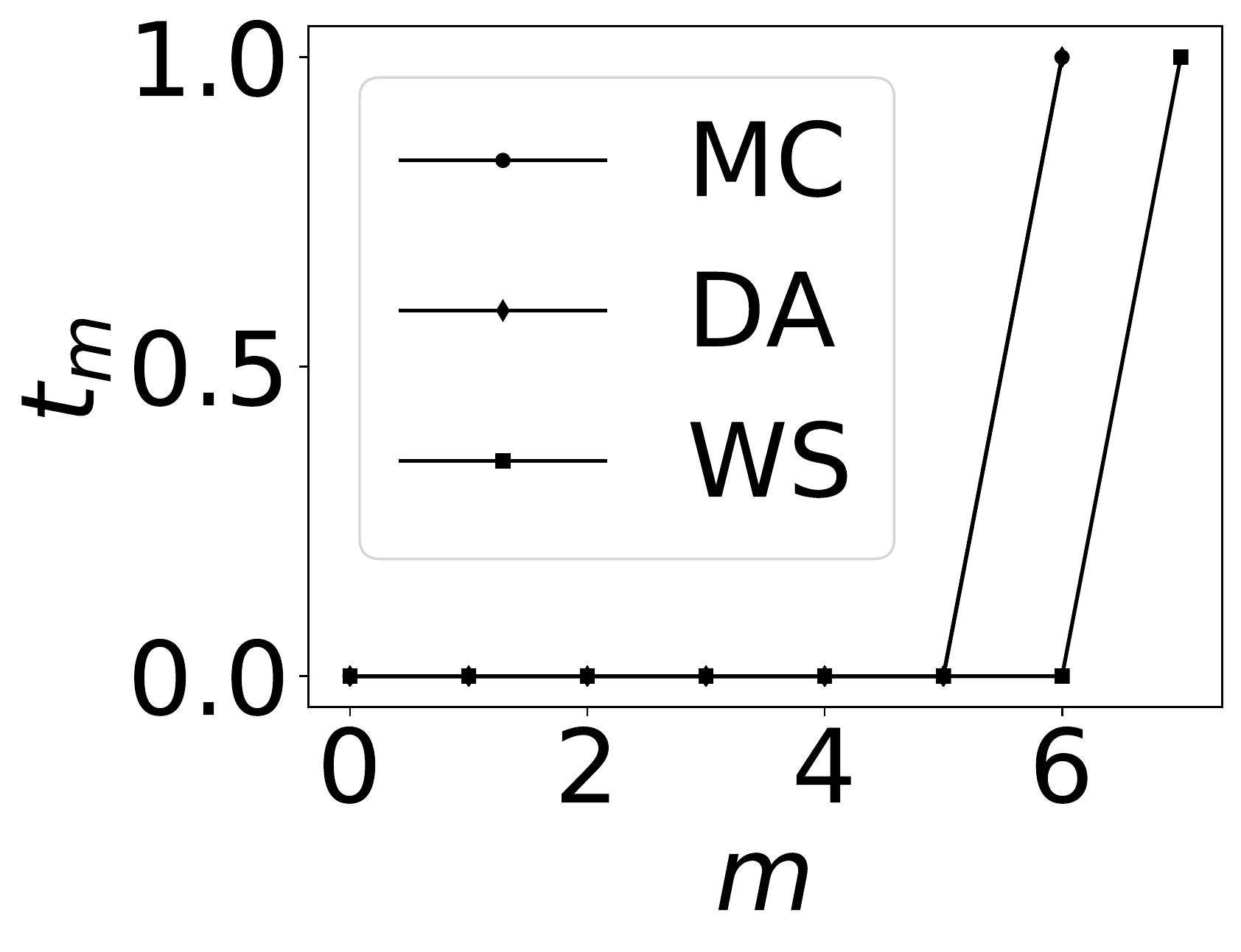}}
\subfloat[]{\includegraphics[width=0.16\textwidth]{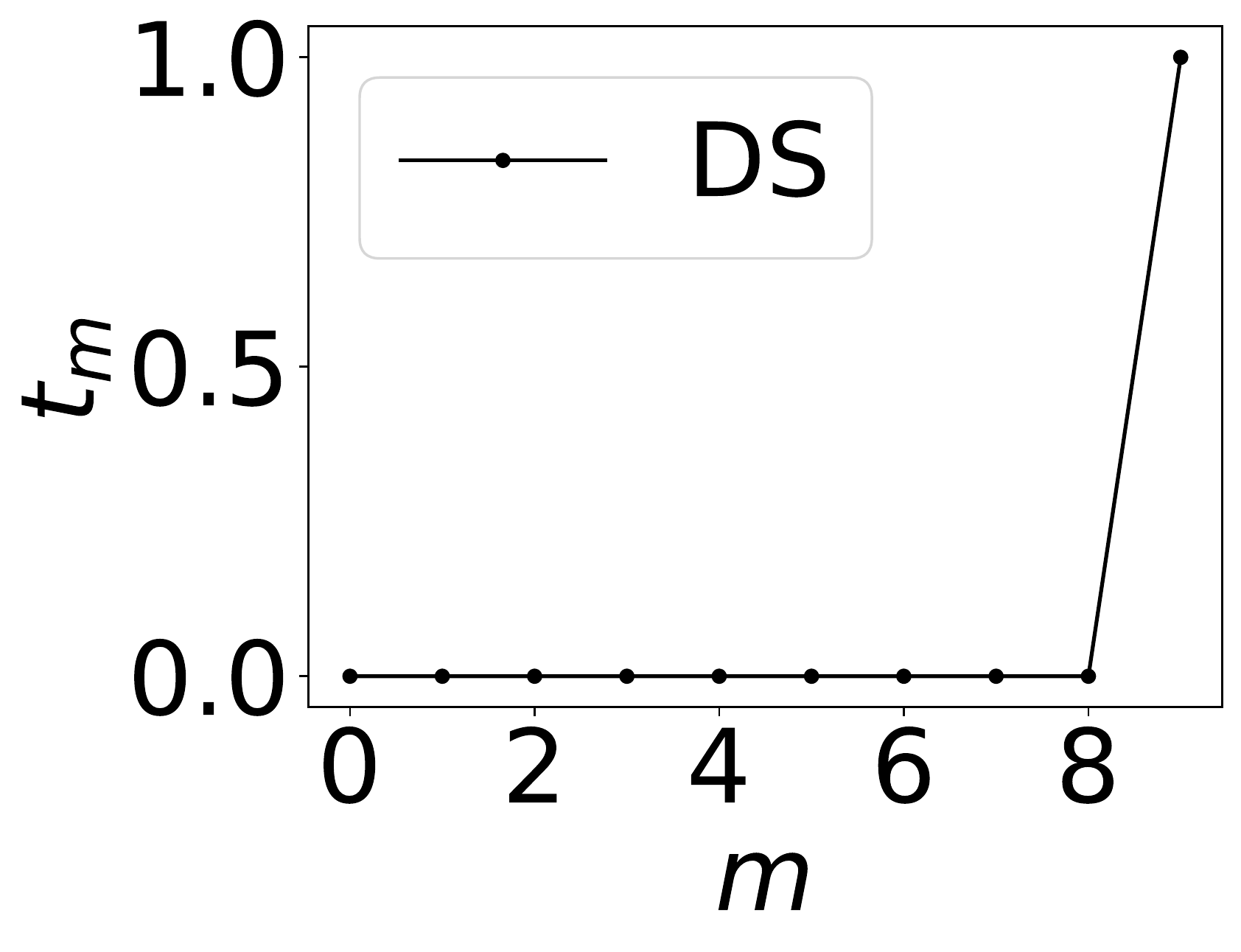}}\\[-6ex]

\subfloat[DNA]{\includegraphics[width=0.16\textwidth]{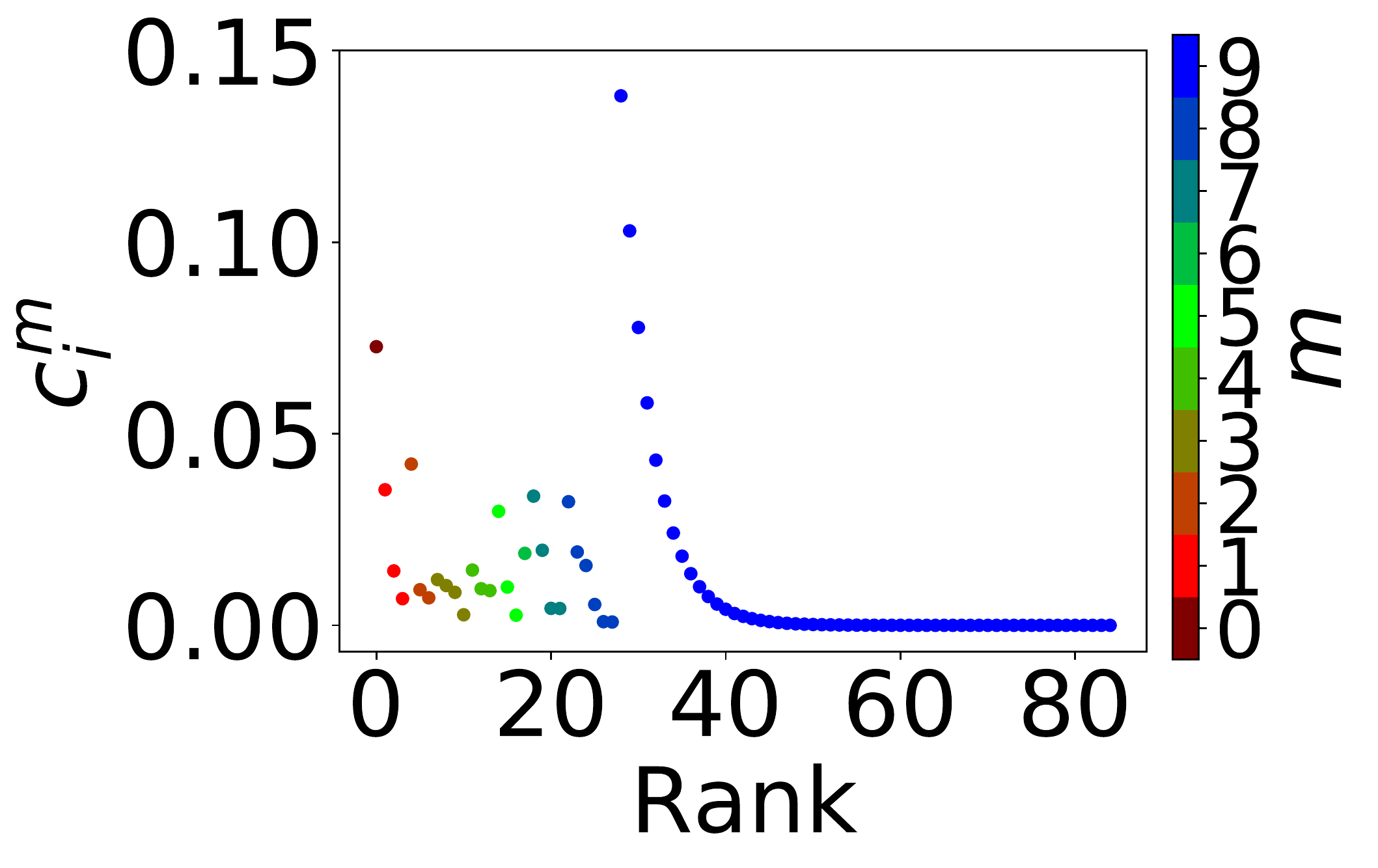}}
\subfloat[Language]{\includegraphics[width=0.16\textwidth]{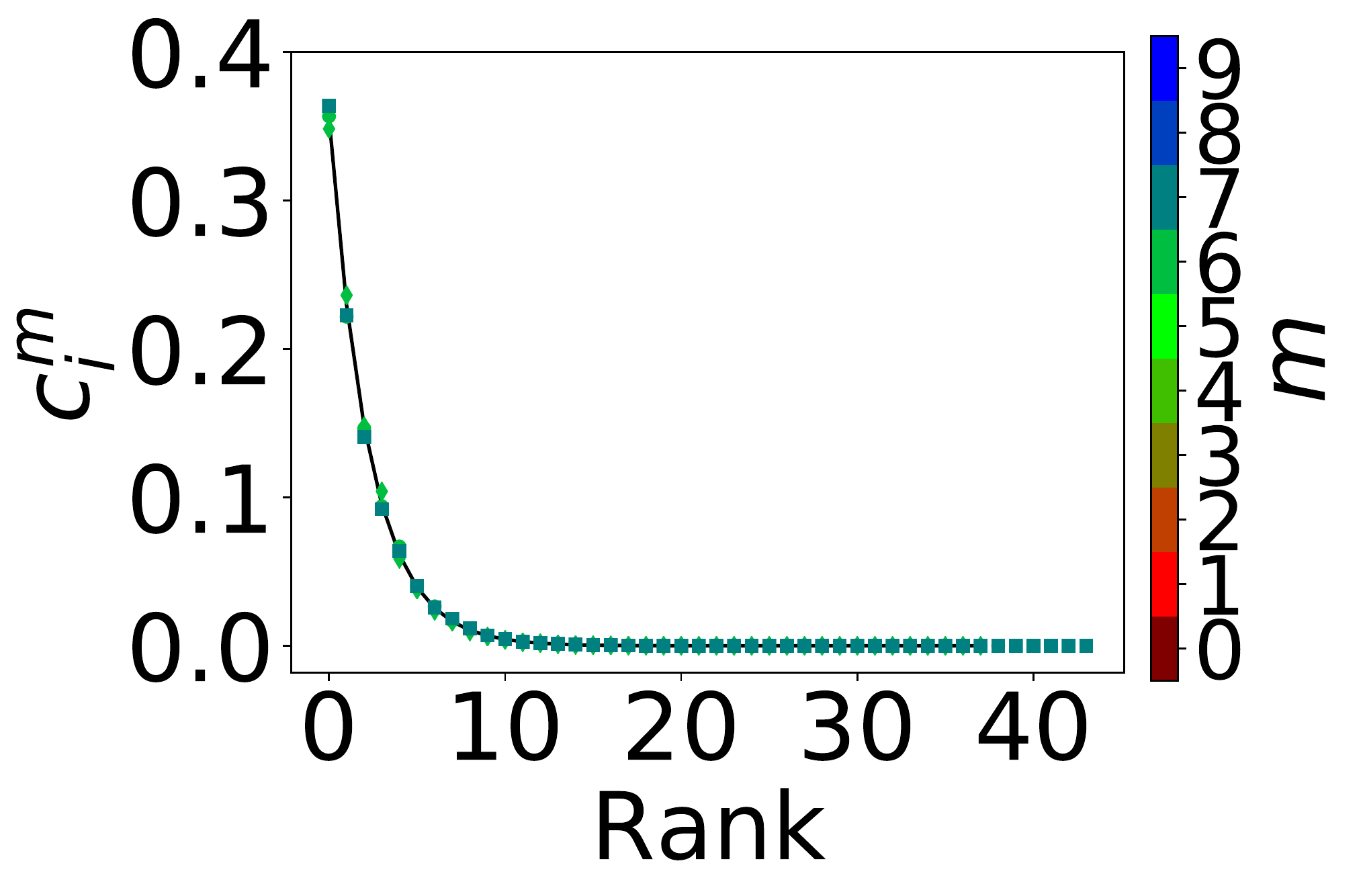}}
\subfloat[Chaos]{\includegraphics[width=0.16\textwidth]{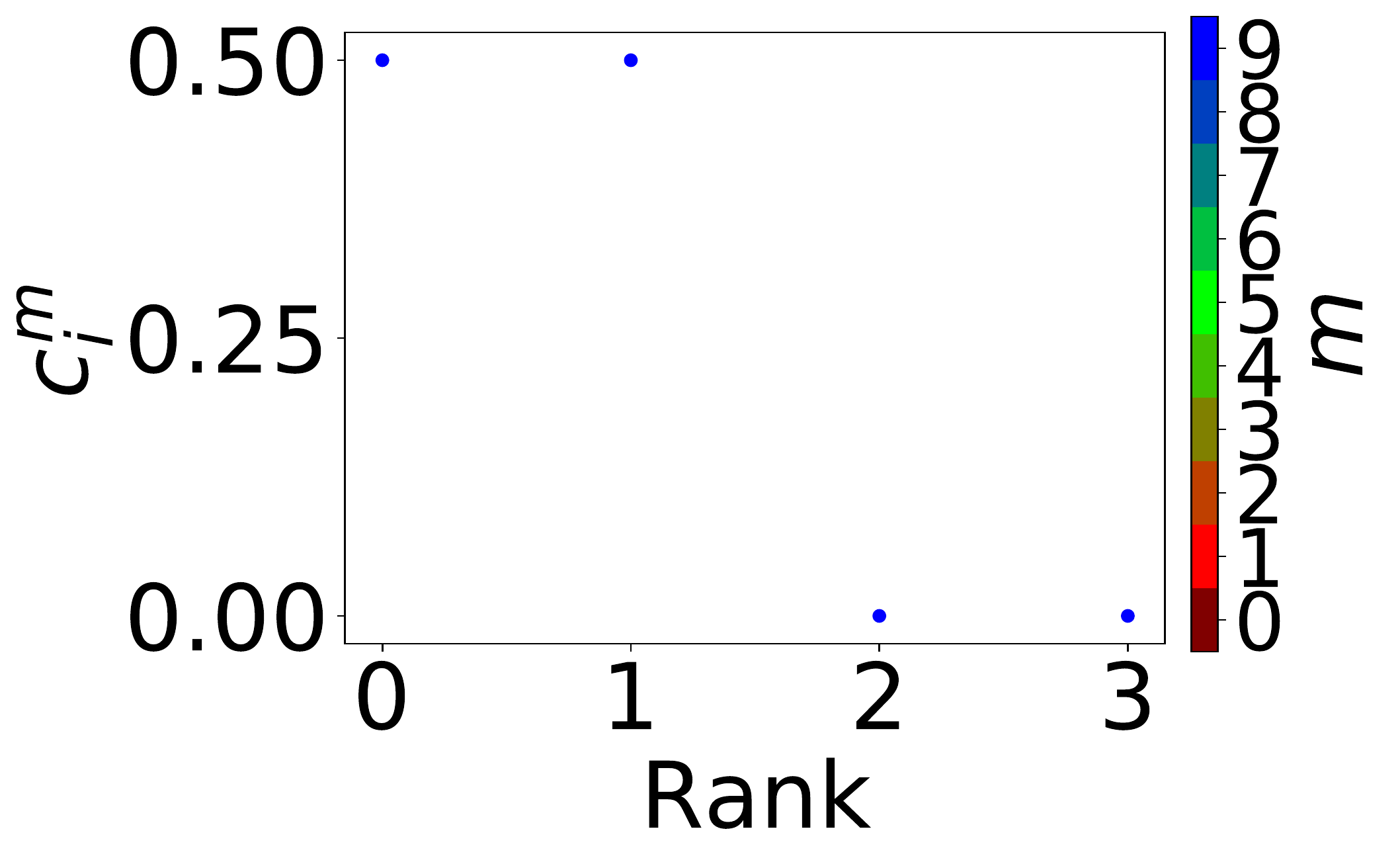}}
\caption{Memory profile $t_m$ (top panels) and matrix decomposition (bottom panels) of real sequences (genetic material, text from literature classics, and the Dragon sequence). In the bottom panels we plot the weights $c^m_i$ of the decomposition in Eq.~(\ref{dek}), sorted in decreasing order for each value of $m$.}
\label{rel}
\end{figure}
 
\subsection{DNA} We have studied the second chromosome of the fruit fly {\it Drosophila melanogaster} (DM) by selecting a DNA sequence of length $1.7 \times 10^7$ from an alphabet ${\cal A}=\{ A,C,T,G\}$ of four letters. The first column of Fig. \ref{rel} shows that even if the estimated order is $M_T=9$, an important fraction of the information of the higher-order Markov chain is contained in subprocesses of lower orders. Therefore, roughly half of the correlations that one would associate to statistics at $m=9$ are spurious and can be reduced. DNA is known to exhibit long-range correlations \cite{92peng,94mantegna,95arneodo,95buldyrev,95allegrini}; however, these appear for orders much higher than the maximal order studied here, and therefore testing whether they can be reduced would require a reformulation of the algorithm as described in the Appendix, Sec. \ref{ap:lfs}.

\subsection{Language} 
We have translated three prominent literature classics ({\it Don Quijote de la Mancha} MC, {\it La Divina Commedia} DA and {\it Hamlet} WS) into Morse Code using the alphabet $\mathcal{A}=\{$``.'',``-'',`` ''$\}$
of only three symbols obtaining, respectively, sequences of lengths $8.7 \times 10^6$, $2.4 \times 10^6$, and $7.4 \times 10^5$.
In all three cases the maximal order $M_T$ coincides with the security cut-off and the process is fully dominated by subprocesses of maximal order $m=M_T$, suggesting that the real order could be larger.
Moreover, we have found that the coefficients of the subprocesses follow an exponential probability function, 
which uncovers a ranked organization of the building blocks of language when expressed in Morse code that goes beyond the Zipf's distribution for the frequencies of words \cite{zipf}.

\subsection{Deterministic chaos} 
As a last example we considered the dragon curve (DS), a deterministic process with fractal properties \cite{dragon}.
We have generated sequences of length $L=5.2 \times 10^5$ ($18$ iterations) with an alphabet $\mathcal{A}=\{L,R\}$ of two letters representing the directions in the rotation of the dragon, either left or right.
The decomposition shows that from all the possible processes at $M_T=9$, only four are present, of which two dominate the transition matrix.
This shows how despite the complexity and the number of parameters of higher-order markov chains, some processes may be decomposed with a small number of subprocesses.
This implies that the model has a very low entropy, as it can be compressed in just two numbers, and is able to capture the deterministic nature of the original sequence. 

\section{Conclusion}
In conclusion, we have proposed a method to represent the mechanism generating a sequence of symbols as a mixture of processes of well-defined orders.
This enables one to determine the memory profile of the underlying Markov process, which is an efficient way of characterizing the causal relations hidden in the sequence.
We hope our method will become a standard tool in the analysis of high-order Markov chains.

\section*{Acknowledgments}
U.A.-R. acknowledges support from the Spanish Government through Maria de Maeztu excellence accreditation 2018-2022 (Ref. MDM-2017-0714),
from the Basque Government through the Posdoctoral Program (Ref. POS-2017-1-0022) and from the Swiss National Science Foundation (Ref. 176938).
V. L. acknowledges support from the EPSRC project EP/N013492/1 and from the Leverhulme Trust Research Fellowship ``CREATE: The network components of creativity and success''.

\clearpage

\appendix
\section*{Appendix}
\subsection{Matrix notation}
\label{ap:mn}
As a simple case to illustrate our notation, let us consider a
sequence $S$ of symbols from an alphabet
with $A=2$ and with symbols $\{0,1\}$. We first need to construct 
matrices $T^m$ with $m=0,1,\ldots$ from the transition probabilities
in Eq. \eqref{tp} in the main text. Suppose the matrix for $m=3$ reads
\begin{equation}
T^3=\left( \begin{array}{cccc} 0.1 & 0.8 & 0.3 & 0.6 \\ 0.9 & 0.2 & 0.7 & 0.4 \end{array} \right).
\label{T}
\end{equation}
This means that $\pi(0|00)=0.1$, $\pi(1|00)=0.9$, $\pi(0|01)=0.8$, $\pi(1|01)=0.2$, $\pi(0|10)=0.3$, $\pi(1|10)=0.7$, $\pi(0|11)=0.6$, and $\pi(1|11)=0.4$.\\

\subsection{Statistical distance}
\label{ap:sd}
Let us see how $\sigma$ is equivalent to the normalized norm of the difference vector.
\begin{eqnarray}
\nonumber d&=&\frac{1}{2} \sum^{D}  |x_i - y_i| = \frac{1}{2} \sum^{D} \max(x_i,y_i) - \min(x_i,y_i) \\ \nonumber &=& \frac{1}{2} \sum^{D} x_i + y_i - 2\min(x_i,y_i) \\ \nonumber  &=& \frac{1}{2} \left( \sum^D x_i + \sum^D y_i -2 \sum^D \min(x_i,y_i) \right) \\ \nonumber &=& \frac{1}{2} (1+1-2\sum^D \min(x_i,y_i)) = 1 - \sum^D \min(x_i,y_i) \\ &=& 1-\sigma .
\end{eqnarray}
Up to this point, it seems unnecessary to make use of an alternative definition, if this is equivalent to the standard one. The reason supporting our decision is clarified when working with more than two distributions. If we add a new one, $z$, the overlap or intersection is calculated as
\begin{equation}
\sigma(x,y,z) = \sum^D_i \min(x_i,y_i,z_i) . 
\end{equation}
The same can be done employing the normalized vector distance, but not in such a simple manner.\\ 

\subsection{Natural label}
\label{ap:nl}
Let us see how the mapping works in an example with $A=2$, $m=3$, and $n=9$. The idea is to retrieve $n$ from the matrix expression.
\begin{equation}
C^3_9=\left( \begin{array}{cccc} 0&1&1&0\\1&0&0&1 \end{array} \right) .
\label{pes}
\end{equation}
As introduced in the main text, the natural label formula is given by
\begin{equation}
n^m_i=\sum^{A-1}_{\alpha=0} \sum^{A^{m-1}-1}_{\beta=0} e_{\alpha\beta} \alpha A^{\beta} , 
\label{nat}
\end{equation}
where $e_{\alpha \beta}$ are the elements of $C^m_i$. Since $C^m_i$ are stochastic Boolean, there is a single non-zero element per column, and therefore, the first summation can be reduced to the rows $\alpha$, such that $e_{\alpha \beta}=1$. Following this expression we have
\begin{equation}
n^3= 1 \times 2^0 + 0 \times 2^1 + 0 \times 2^2 + 1 \times 2^3 = 9 .
\end{equation}

\subsection{Number reduction and extension mechanism}
\label{ap:nre}
Let us first consider the extension of the matrix of the previous section in Eq. \eqref{pes}:
\begin{equation}
C^{3[+]1}_9=\left( \begin{array}{cccccccc} 0&1&1&0&0&1&1&0\\1&0&0&1&1&0&0&1 \end{array} \right) .
\end{equation}
The associated number $n^{3[+]1}_9 =153$, is computed by either using Eq. \eqref{nat} from the Appendix or Eq. \eqref{natex} from the main text. Since the first option has already been explained in the previous section we go for the second one. 
\begin{equation}
n^{3[+]1}_{9}=n^{3}_{9}\frac{2^{2^{3}}-1}{2^{2^{3-1}}-1}=n^{3}_{9}\times 17 = 153.
\end{equation}

Now that we have explored the number extension, we try the opposite, the number reduction mechanism. We are interested in knowing if $C^4_{153}$ has $m=4$ as its true order. In order to test that, we have to try the divisibility of $153$ with $17$ since $\frac{2^{2^{4-1}}-1}{2^{2^{4-2}}-1}=17$. We get the expected result, $153=17 \times 9$. Our matrix does not belong to order $m=4$, and the label of our matrix in order $m=3$ is $n^{3}_9=9$, as we obviously knew because that has been our starting point.\\ 

We try once more and see if the same process can be expressed in order $m=2$. In order to do so we have to test the divisibility of $9$ with $5$, since $\frac{2^{2^{3-1}}-1}{2^{2^{3-2}}-1}=5$. The division does not retrieve a natural number, so the true order of the process is $m=3$.\\ 

\subsection{Decomposition algorithm}
\label{ap:da}
Let us now show how to decompose matrix $T^3$ given in Eq.~\eqref{T} as
in Eq. \eqref{dek} of the main text.  
We begin from the term corresponding to $m=0$, using the equal probabilities
$1/A$ of the uniform model and the reduced matrix in Eq. \eqref{eqRm} of the main text. We first get $R^{0}=\min T^2_{\alpha \beta} = 0.1$ and $c^{0}_{0}= A \times R^{0} = 0.2$, with
$C^{0}_{0} = 0.5$ corresponding to the uniform model. Since the extension of $C^{0}_{0}$ is
\begin{equation}
C^{0[+]2}_0 = \left( \begin{array}{cccc} 0.5 & 0.5 & 0.5 & 0.5 \\ 0.5 & 0.5 & 0.5 & 0.5 \end{array} \right),
\end{equation}
we can then substract the first term of the decomposition: 
\begin{eqnarray}
&& T^3-c^{0}_0 C^{0[+]2}_0  \\
\nonumber &&=\left( \begin{array}{cccc} 0.1 & 0.8 & 0.3 & 0.6 \\ 0.9 & 0.2 & 0.7 & 0.4 \end{array} \right) - \left( \begin{array}{cccc} 0.1 & 0.1 & 0.1 & 0.1 \\ 0.1 & 0.1 & 0.1 & 0.1 \end{array} \right) \\
\nonumber &&= \left( \begin{array}{cccc} 0 & 0.7 & 0.2 & 0.5 \\ 0.8 & 0.1 & 0.6 & 0.3 \end{array} \right).
\end{eqnarray}
Since $C^{0}_0$ is the only matrix at $m=0$, there is no need to search for more compatible ones. In any case, the new reduced matrix is $R^{0}=0$, so we jump to the next level. We can move on to construct the contribution due to $m=1$.  In the first cycle of $m=1$, $R^1$ is
\begin{equation}
R^1=\left( \begin{array}{c} 0\\0.1 \end{array} \right).
\end{equation}
Therefore, we obtain $c^1_1=\max \{0,0.1\}=0.1$, and $C^1_1= \left( \begin{array}{c} 0 \\ 1 \end{array} \right)$. Since the extension of $C^1_1$ is
\begin{equation}
C^{1[+]2}_1 =  \left( \begin{array}{cccc} 0 & 0 & 0 & 0 \\ 1 & 1 & 1 & 1 \end{array} \right),
\end{equation}
we can get the resultant matrix when we have removed 
\begin{eqnarray}
&& T^3 -c^{0}_0 C^{0[+]2}_0 -c^1_1 C^{1[+]2}_1 \\
\nonumber &&=\left( \begin{array}{cccc} 0 & 0.7 & 0.2 & 0.5 \\ 0.8 & 0.1 & 0.6 & 0.3 \end{array} \right) - \left( \begin{array}{cccc} 0 & 0 & 0 & 0 \\ 0.1 & 0.1 & 0.1 & 0.1 \end{array} \right) \\
\nonumber  &&= \left( \begin{array}{cccc} 0 & 0.7 & 0.2 & 0.5 \\ 0.7 & 0 & 0.5 & 0.2 \end{array} \right).
\end{eqnarray}
If we compute $R^1$ we will see that is null, so we can jump to the next level. We have 
\begin{equation}
R^2=\left( \begin{array}{cc} 0&0.5\\0.5&0 \end{array} \right),
\end{equation}
which means that $c^2_1=\min \{0.5,0.5\}=0.5$. The matrix $C^2_1$ and its extension $C^{2[+]1}_1$ read
\begin{equation}
C^2_1=\left( \begin{array}{cc} 0&1\\1&0 \end{array} \right), \hspace{0.2cm} C^{2[+]1}_1=\left( \begin{array}{cccc} 0 & 1 & 0 & 1 \\ 1 & 0 & 1 & 0 \end{array} \right).
\end{equation}
In terms of these, we calculate the resultant total matrix.
\begin{eqnarray}
&& T^3 -c^{0}_0 C^{0[+]2}_0 -c^1_1 C^{1[+]2}_1 - c^2_1 C^{2[+]1}_1 \\
\nonumber &&=  \left( \begin{array}{cccc} 0 & 0.7 & 0.2 & 0.5 \\ 0.7 & 0 & 0.5 & 0.2 \end{array} \right) -  \left( \begin{array}{cccc} 0 & 0.5 & 0 & 0.5 \\ 0.5 & 0 & 0.5 & 0 \end{array} \right) \\
\nonumber &&= \left( \begin{array}{cccc} 0 & 0.2 & 0.2 & 0 \\ 0.2 & 0 & 0 & 0.2 \end{array} \right) .
\end{eqnarray}
Again we have a null column in $R^2$, so we can jump to the next and last level. In this case no calculations are needed, since the remaining matrix can be expressed as a Boolean matrix, namely, $C^3_9$, multiplied by a constant, $c^3_9=0.2$. No extension is needed in this time since the order of $C^3_9$, $m=3$, is already $M_T$.
\begin{equation}
C^3_9=\left( \begin{array}{cccc} 0 & 1 & 1 & 0 \\ 1 & 0 & 0 & 1 \end{array} \right) .
\end{equation}
After the complete process we have
\begin{equation}
T^3=0.2 C^{0[+]2}_0 +0.1 C^{1[+]2}_1 + 0.5 C^{2[+]1}_1 + 0.2 C^3_9
\end{equation}
and more explicitly
\begin{eqnarray}
&&\left( \begin{array}{cccc} 0.1 & 0.8 & 0.3 & 0.6 \\ 0.9 & 0.2 & 0.7 & 0.4 \end{array} \right) = \\
\nonumber &&\left( \begin{array}{cccc} 0.1 & 0.1 & 0.1 & 0.1 \\ 0.1 & 0.1 & 0.1 & 0.1 \end{array} \right) + \left( \begin{array}{cccc} 0 & 0 & 0 & 0 \\ 0.1 & 0.1 & 0.1 & 0.1 \end{array} \right) + \\
\nonumber &&\left( \begin{array}{cccc} 0 & 0.5 & 0 & 0.5 \\ 0.5 & 0 & 0.5 & 0 \end{array} \right)+\left( \begin{array}{cccc} 0 & 0.2 & 0.2 & 0 \\ 0.2 & 0 & 0 & 0.2 \end{array} \right) 
\end{eqnarray}

\subsection{Low frequency strings}
\label{ap:lfs}
We introduce an upper bound in $m$ to make sure that the frequencies of the strings involved in the calculation of the transition probabilities are high enough. We impose a first cut-off at $m+2 \le\log_A L$ even before reading the values of S. This cut-off implies that the average string frequency is $f(x^{m+2})=1$ in a uniform model. A second threshold is introduced after reading the string frequencies in $S$: when a string of length $m+1$ is unique, $f(x^{m+1})=1$, we impose an upper bound at $m+2$. In practice this means that one cannot extract higher-order correlations from sequences in which they are potentially present. In order to dodge this drawback, one can extract the transition probabilities from ensembles of sequences, always under the assumption that all the samples have been produced by the same Markov process. These restrictions are not needed for running the decomposition algorithm, however we still need to provide the transition probabilities for strings that are not found in $S$. In order to do so, we employ a simple smoothing technique for strings of null frequency: we compute their transition probabilities $\pi$ by copying the ones of the previous order, which is equivalent to extend the transition matrix for the columns corresponding to those null frequency strings.\\

\end{document}